\documentstyle[12pt]{article}

\newcommand{\GeV}{\hbox{GeV}}
\newcommand{\MeV}{\hbox{MeV}}
\newcommand{\bbf}{\em}

\newcommand{\draft}{
\newcount\timecount
\newcount\hours \newcount\minutes  \newcount\temp \newcount\pmhours

\hours = \time
\divide\hours by 60
\temp = \hours
\multiply\temp by 60
\minutes = \time
\advance\minutes by -\temp
\def\hour{\the\hours}
\def\minute{\ifnum\minutes<10 0\the\minutes
            \else\the\minutes\fi}
\def\clock{
\ifnum\hours=0 12:\minute\ AM
\else\ifnum\hours<12 \hour:\minute\ AM
      \else\ifnum\hours=12 12:\minute\ PM
            \else\ifnum\hours>12
                 \pmhours=\hours
                 \advance\pmhours by -12
                 \the\pmhours:\minute\ PM
                 \fi
            \fi
      \fi
\fi
}
\def\fullclock{\hour:\minute}
} 
\newcommand{\singlecolumn}{
\advance\textheight by 60pt
\advance\voffset by -40pt
\advance\textwidth by 50pt
\advance\oddsidemargin by -25pt
\advance\evensidemargin by -25pt
\def\baselinestretch{1}
\def\single_space{\baselineskip 12pt plus 1pt minus 1pt}
\def\one_and_a_half_space{\baselineskip 19pt plus 1pt minus 1pt}
\def\double_spacesp{\baselineskip 25pt plus 2pt minus 2pt}
} 

\singlecolumn

\begin{document}
\newcommand{\lfs}{\hspace{-0.11in}/}
\newcommand{\lfz}{\hspace{-0.09in}/}
\newcommand{\geqnew}{\stackrel{>}{\!\ _{\sim}}}
\newcommand{\leqnew}{\stackrel{<}{\!\ _{\sim}}}
\renewcommand{\thesection}{\Roman{section}}

\begin{titlepage}
\draft
\begin{flushright}
{\bf
PSU/TH/135\\
TAUP 2099-93\\
hep-ph/9310346\\
October 1993\\
}
\end{flushright}
\vskip 1.cm
{\Large
{\bf
\begin{center}
Aspects of heavy quark production\\
in polarized proton-proton collisions
\end{center}
}}
\vskip 0.8cm
\begin{center}
M. Karliner\\
School of Physics and Astronomy
\\ Raymond and Beverly Sackler Faculty of Exact Sciences
\\ Tel-Aviv University, 69978 Tel-Aviv, Israel
\\ e-mail: marek@vm.tau.ac.il
\vskip 0.4cm
and
\vskip 0.4cm
R. W. Robinett\\
Department of Physics\\
The Pennsylvania State University\\
University Park, PA 16802 USA\\
e-mail: rq9@psuvm.bitnet
\end{center}
\vskip 1.0cm
\begin{abstract}

We examine the spin-dependence of $b$-quark production at large
transverse momentum in polarized proton-proton collisions. The
leading-order (LO) $2 \rightarrow 2$ subprocesses, $g + g \rightarrow
Q + \overline{Q}$ and $q + \overline{q} \rightarrow Q + \overline{Q}$
have a large `analyzing power', as measured by the
average spin-spin asymmetry $<\!\!\hat{a}_{LL}\!\!>$,
 at large $p_T$, approaching $-100\%$. The contributions
from next-to-leading order (NLO)
$2 \rightarrow 3$ processes are also discussed and found
to be dominated at large transverse momentum
by subprocesses with a large and positive partonic level
spin asymmetry leading to strong cancellations with the LO
predictions. The results suggest that
$b$-quark production might constitute an interesting test of the
spin-dependence of NLO QCD but they also
point up the
importance of a full calculation of the complete $O(\alpha_S^3)$
spin-dependent corrections for a successful extraction of the polarized
gluon distribution. Similarities to gluino pair production are
also briefly discussed.

\end{abstract}
\end{titlepage}
\double_spacesp

\section{Introduction}

The first results on polarized deep-inelastic scattering from the EMC
experiment \cite{EMC} were interpreted \cite{EMC,EFR} as evidence that
the total spin of the proton carried by quarks was rather small, in
fact compatible with zero. This suggested that the  polarization of
sea quarks, including non-valence strange quarks, might be larger than
naively expected from the quark model.
An  alternative   explanation  had  also   been  suggested
\cite{DeltagI,DeltagII,DeltagIII}, based on the contribution from gluon
polarization to the matrix element of the flavor-singlet axial current.
There are strong arguments \cite{EK,Box} that the glue polarization
is unlikely to be as large as needed to explain the EMC data, but the
ultimate resolution must come from experiment.
The more recent, next generation SMC $e-$ deuterium \cite{NMC} and
SLAC $\mu-$ $^3$He \cite{SLAC} experiments have been recently analyzed
\cite{BjSR,CR,PANIC} and found to be compatible with each other and with
the original EMC data. The possibility of a large gluon polarization
therefore continues to be of interest.

The lack of the analog of polarized charged-current scattering implies
that far less flavor separation is possible when examining spin-dependent
parton distributions.\footnote{Polarized charged-current scattering
could, in principle, be realized with a neutrino beam and a polarized
nucleon target, but the very small
cross-section is likely to make such an experiment
impractical.}
 The integral constraint on the polarized
parton distributions arising from the spin-1/2 nature of the proton
includes possible contributions from constituent parton orbital
angular momentum and so is far less restrictive than the corresponding
momentum sum rule.
All of these observations have pointed to the desirability of
more direct measurements
of the polarized gluon distribution in polarized hadron collisions
akin to the information available on the unpolarized gluon distribution
from, say,  direct photon production.

At the same time, the prospects \cite{workshop,particleworld}
for the availability of high luminosity
polarized proton-proton collisions at the Relativistic Heavy Ion
Collider (RHIC) at collider energies ($\sqrt{s}=50-500\,\GeV$) have been
extensively discussed.  These facts have
motivated  an increasingly large literature discussing
the spin-dependence
and sensitivity to polarized parton distributions of many standard model
processes.  Familiar processes such as jet production \cite{jets,jets2},
direct photons \cite{photon}, and weak boson production \cite{weakboson}
have all been studied as have processes such as quarkonium production
at both low $p_T$ \cite{cortes}--\cite{mike} and high $p_T$
\cite{rick2,doncheskikim} and double photon production \cite{2gamma}.
 Many of the results of these studies have found their
way into the proposals \cite{proposal}
for spin-physics experiments using the
proposed RHIC detectors.  Using such facilities, a comprehensive program
involving the extraction of the polarized parton densities and tests
of the spin-structure of the standard model is envisaged.

The availability of high quality data on $b$-quark production from both
UA1 \cite{UA1b} and CDF \cite{CDFb} combined with the availability of
a full set of next-to-leading (NLO) order QCD calculations of heavy quark
cross-sections \cite{dawson1,dawson2,smith} have proved
to be a valuable new
constraint on unpolarized gluon distributions \cite{meng,mangano}.
 It is thus
natural to explore the spin-dependence of $b$-quark production
and its sensitivity to the polarized gluon density in
polarized proton-proton collisions at RHIC energies.
In general, one expects that theoretical estimates for
$b$-quark production are more reliable than for charm
production. In the following we will therefore concentrate
on $b$-quarks.

Heavy quark production in polarized $pp$ collisions
was first examined by Contogouris {\it et al.} \cite{conto} who
considered only the spin-dependence of the lowest-order
$2 \rightarrow 2$ processes
$g + g, q +\overline{q} \rightarrow Q + \overline{Q}$ and their impact
on the total $b$-quark production cross-section. (An earlier
study by Hidaka \cite{hidaka} derives some of the same helicity-dependent
cross-sections as in Ref.~\cite{conto} for use in the study of
spin-dependence in charmonium production. Photoproduction of
charm via the similar leading-order
$\gamma + g \rightarrow Q + \overline{Q}$ process
has recently been examined in Ref. \cite{keller}.)

In this report, we extend this analysis and discuss two important
aspects of the problem: $p_T$ dependence and NLO effects.
In Sec. II we analyze the spin-dependence of the $p_T$-dependent
cross-section using the $2  \rightarrow 2$ processes.  This is
motivated by the fact that the existing data extend over a wide range
of transverse momentum.  We find a dramatic dependence of the
observable spin-spin asymmetries on $p_T$.
In Sec.  III we discuss the likely consequences of next-to-leading
order (NLO)  effects.  These  are both vertex  and box-diagram
corrections to the $2 \rightarrow 2$ processes, as well as the $2
\rightarrow 3$ contributions. We find that a large cancellation
between the spin-dependence of the two types of processes is likely at
large $p_T$.  This may make $b$-quark production a less sensitive
probe of the polarized gluon density than naively imagined on the
basis of the leading-order predictions.
It might provide, however,
an important test of the spin-dependent matrix-elements of QCD beyond
the leading order.  We do not attempt in this study any
systematic analysis of the
detectability of heavy quark production in the proposed RHIC detectors
nor make any quantitative estimates of the sensitivity of such
experiments to the extraction of the polarized gluon distributions.

\section{$2 \rightarrow 2$ processes}

We begin by discussing the spin-dependence of the leading-order
$ 2 \rightarrow 2$ subprocesses, $q +\overline{q} \rightarrow
Q +\overline{Q}$ and $g + g \rightarrow Q + \overline{Q}$.  The
cross-section for $q(q_1) + \overline{q}(q_2)
\rightarrow Q(k_1) + \overline{Q}(k_2)$ can be written as
\begin{equation}
\frac{d\hat{\sigma}}{dt} = \frac{4 \pi \alpha_S^2}{9 \hat{s}^2}
\left[\frac{\tilde{t}^2 + \tilde{u}^2 + 2M^2\hat{s}}{\hat{s}^2} \right]
\end{equation}
where
\begin{equation}
\tilde{t} \equiv M^2 - \hat{t} \;\;\;\;\;\;\;\;,\;\;\;\;\;\;\;
\tilde{u} \equiv M^2 - \hat{u},
\end{equation}
\begin{equation}
\hat{s} = (q_1 + q_2)^2 \;\;\;,\;\;\; \hat{t} = (q_1-k_1)^2\;\;\;,\;\;\;
\hat{u} = (q_1 - k_2)^2
\end{equation}
and $M$ is the heavy quark mass.
The partonic level longitudinal spin-spin asymmetry is defined by
\begin{equation}
\hat{a}_{LL} \equiv \frac{d\hat{\sigma}(++) - d\hat{\sigma}(+-)}
{d\hat{\sigma}(++) + d\hat{\sigma}(+-)}
\end{equation}
where $\pm$ refers to the helicity of the incoming partons.  In this
case, it is easily seen that
\begin{equation}
\hat{a}_{LL}(q\overline{q} \rightarrow Q \overline{Q}) = -1
\end{equation}
 due to helicity
conservation and the assumed masslessness of the initial state quarks.

For the subprocess $g(q_1) + g(q_2) \rightarrow
Q(k_1) + \overline{Q}(k_2)$, the unpolarized
cross-section, i.e. averaged (summed)
 over initial (final) spins can be written in the factorized
form \cite{sexton}
\begin{eqnarray}
\frac{d \hat{\sigma}}{dt} & = & \frac{\pi \alpha_S^2}{8 \hat{s}^2}
\left(\frac{4}{3\tilde{u}\tilde{t}} - \frac{3}{\hat{s}^2} \right)
\left[(\tilde{u}^2 + \tilde{t}^2) +
\left(\frac{4M^2 \hat{s}}{\tilde{t}\tilde{u}} \right)
(\tilde{u}\tilde{t} - M^2 \hat{s}) \right] \nonumber \\
& = & \frac{1}{2}\left( \frac{d \hat{\sigma}(++)}{d t}
+ \frac{ d\hat{\sigma}(+-)}{dt} \right)
\end{eqnarray}
where $\tilde{t},\tilde{u}$ are defined above.  A simple calculation
using FORM \cite{form} gives for the difference in helicity-dependent
cross-sections
\begin{equation}
\frac{d\hat{\sigma}(++)}{dt} - \frac{d\hat{\sigma}(+-)}{dt}
= -\frac{\pi \alpha_S^2}{4\hat{s}^2}
\left(\frac{4}{3\tilde{u}\tilde{t}} - \frac{3}{s^2} \right)
\left[(\tilde{u}^2 + \tilde{t}^2) -
\left(\frac{2M^2\hat{s}}{\tilde{u}\tilde{t}}\right)
( \tilde{u}^2 + \tilde{t}^2)
\right]
\end{equation}
which, after some manipulation, can be shown to be identical to the
equivalent result of Ref.~\cite{conto}.
The corresponding partonic level
spin asymmetry is then simply given by
\begin{equation}
\hat{a}_{LL}(gg \rightarrow Q\overline{Q})=
-\frac{
(\tilde{u}^2 + \tilde{t}^2)
- 2M^2\hat{s}(\tilde{u}^2 + \tilde{t}^2)/\tilde{u}\tilde{t}
}{
(\tilde{u}^2 + \tilde{t}^2)
+ 4M^2\hat{s}(\tilde{u}\tilde{t} - M^2\hat{s})/\tilde{u}\tilde{t}
}\;\;\;\;.
\end{equation}
We note that these expressions for the total and helicity-dependent
cross-sections are quite similar to existing results for gluino
pair-production \cite{renard}
via $g + g \rightarrow \tilde{g} + \tilde{g}$.
The only difference arises in the color-dependent pre-factor which
in the case of color-octet gluino production is
$(1/\tilde{t}\tilde{u} - 1/\hat{s}^2)$.
The resulting partonic level asymmetry is  then identical to Eqn. 8.
This similarity is not surprising as both involve the
pair-production of colored, spin-1/2 fermions via gluon fusion.

Using the kinematic relations
\begin{equation}
\tilde{t} = \frac{\hat{s}}{2}\left(1-\beta y\right) \;\;\;,\;\;\;
\tilde{u} = \frac{\hat{s}}{2}\left(1+\beta y\right)
\end{equation}
where $y \equiv \cos(\theta^*)$ gives the center-of-mass scattering
angle and $\beta \equiv \sqrt{1-4M^2/\hat{s}}$ is the heavy quark speed,
we
plot $\hat{a}_{LL}$ versus $y$ for several values of
$\sqrt{\hat{s}}/2M$ in Fig. 1(a).  We note that very near threshold, i.e.
$\sqrt{\hat{s}} \geqnew 2M$ the asymmetry is $+1$, independent of $y$,
while
for large energies it reduces to the massless quark limit
$\hat{a}_{LL} = -1$, first found by Babcock, Monsay, and Sivers
\cite{sivers}.  For comparison,
we plot in Fig. 1(b) the corresponding $\hat{a}_{LL}$
for the $g + g \rightarrow g + g$ and $q + g \rightarrow q + g (\gamma)$
subprocesses which give important contributions to jet and direct
photon production.

For the case of massive final state quarks, the individual
helicity-dependent differential cross-sections can be integrated,
so we can
define a partonic level spin-spin asymmetry for the total cross-section
via
\begin{equation}
\hat{A}_{LL} \equiv \frac{\hat{\sigma}(++) - \hat{\sigma}(+-)}
{\hat{\sigma}(++) + \hat{\sigma}(+-)}.
\end{equation}
Integrating Eqns. 6 and 7 we find that
\begin{equation}
\hat{A}_{LL}(\hat{s}) = \frac{2(9\beta^2-17)\log\left( \frac{1+\beta}
{1-\beta}\right)+ 15\beta(5-\beta^2)}
{(33 -18\beta^2 + \beta^4)\log\left( \frac{1+\beta}{1-\beta}\right) +
 \beta(31\beta^2-59) }\;\;\;.
\end{equation}
We plot this asymmetry in Fig. 2 as a function of $\sqrt{\hat{s}}/2M$
and note again the (slow) change from $+1$ at threshold to its asymptotic
value of $-1$.

The observable longitudinal spin-spin asymmetry is given by
\begin{equation}
A_{LL} \equiv
\frac{ \sum_{i,j} \int dx_a \int dx_b
[\Delta f_i(x_a,Q^2) \Delta f_j(x_b,Q^2)]
\hat{a}^{ij}_{LL} d\hat{\sigma}_{ij} }
{ \sum_{i,j} \int dx_a \int dx_b [f_i(x_a,Q^2) f_j(x_b,Q^2)]
d\hat{\sigma}_{ij}}
\end{equation}
where the $\hat{a}_{LL}^{ij}$ and $d\hat{\sigma}_{ij}$ are the
relevant partonic level asymmetries and hard-scattering cross-sections
respectively, summed over all possible initial parton states, $(i,j)$.
  The information
on the spin-dependent parton distributions is contained in the
$\Delta f(x,Q^2)$, defined via
$\Delta f(x,Q^2) \equiv f_{+}(x,Q^2) - f_{-}(x,Q^2)$
where $f_{+}\,(f_{-})$ denotes the parton distribution in a polarized
nucleon with helicity parallel (antiparallel) to the parent nucleon
helicity.

A useful measure of the intrinsic spin-dependence in an observable
process is the average value of the partonic level spin-spin asymmetry,
defined via
\begin{eqnarray}
<\hat{a}_{LL}> & \equiv  &
\frac{ \sum_{i,j} \int dx_a \int dx_b [f_i(x_a,Q^2) f_j(x_b,Q^2)]
\hat{a}^{ij}_{LL} d\hat{\sigma}_{ij} }
{ \sum_{i,j} \int dx_a \int dx_b [f_i(x_a,Q^2) f_j(x_b,Q^2)]
d\hat{\sigma}_{ij}} \nonumber \\
& = & \sum_{ij} <\hat{a}^{ij}_{LL}>
\end{eqnarray}
which can be equated with an `average analyzing power' for the reaction
in question.  Each of the individual `average asymmetries'
\begin{equation}
<\!\hat{a}_{LL}^{ij}\!> \equiv
\frac{\int dx_a \int dx_b [f_i(x_a,Q^2) f_j(x_b,Q^2)]
\hat{a}^{ij}_{LL} d\hat{\sigma}_{ij} }{\sigma_{tot}}
\end{equation}
is both a measure of the spin-dependence (via the $\hat{a}_{LL}$)
of a specific contributing
initial parton configuration and of its importance in the cross-section.
This quantity is a useful figure of merit for the
examination of the spin-dependence in any specific reaction.  The
observable asymmetries, $A_{LL}$, which require information about the
(presently unknown) $\Delta f_i(x,Q^2)$, will of course be smaller than
this quantity, barring accidental cancellations,
since $|\Delta f(x,Q^2)/f(x,Q^2)|<1$.
Given a set of putative polarized distributions, the observable $A_{LL}$
can be easily calculated and estimates of the measurability of any
particular asymmetry can be made by using the total cross-section and
luminosity (and appropriate cuts for the realistic detection of the
events) to estimate the total event rates.  In this report, we will
focus on calculations of $<\!\hat{a}_{LL}\!>$ as it provides a
reasonable glimpse into the relative importance of the various
contributions to the observable spin-dependence.

For $b$-quark production
in  $pp$ collisions in the range of energies accessible to RHIC
($\sqrt{s} = 50-500\,\GeV$), we find that when
calculating the {\bbf total cross-section}
the average partonic center of mass energy, $\sqrt{<\hat{s}>}/2M$,
is in the range $1.2 - 1.9$.
This fact and the results shown in Fig.~1(a) and 2 then  imply
that the observable $A_{LL}$, when typical
polarized gluon distributions
(such as those of Ref.~\cite{altarelli})
are included, does indeed yield a not
unreasonably small and {\bbf positive} asymmetry, just as found by
Contogouris {\it et al.} in Ref.~\cite{conto}.
The total cross-section is,
of course, dominated by events at low transverse momentum, i.e. for
values of $\sqrt{\hat{s}}/2M \leqnew 2$ and for large values of
$|y|$, i.e., $|y| \approx 1$.  From Figs. 1(a) and 2 we can see that
as the center-of-mass energy is increased and as $\theta^* \rightarrow
\pi/2$, i.e., when the transverse momentum is increased, the partonic
level spin-spin asymmetries rapidly approach $-1$ for the dominant
gluon fusion process, changing dramatically the expected magnitude and
even sign of the observable asymmetry.

Motivated by the excellent UA1 \cite{UA1b}
and CDF \cite{CDFb} data on $b$-quark production as a function of
transverse momentum, we calculate the average $<\!\hat{a}_{LL}\!>$
mentioned above for b-quark production in $pp$ collisions at the
highest RHIC energy, $\sqrt{s} = 500\,\GeV$, for the integrated
production cross-section, namely $\sigma(p_T > p_T(min))$ versus
$p_T(min)$.  We use the $2 \rightarrow 2$ subprocesses above, a
(relatively) recent set of leading-order (LO)
unpolarized parton distributions
\cite{duke} (as that is all that is required in Eqn. 13)
corresponding to $\Lambda^{1-loop} = 177\,\MeV$.
\footnote{The qualitative results are likely to be insensitive
to the details of the  unpolarized parton distributions,
since the estimates in Eq.~13 should be understood as an
indication of what
can be expected from polarized distributions.}
In addition, we assume $M = m_b = 5\,\GeV$ and use a momentum scale
$Q^2 = (M^2 + p_T^2)/4$ as suggested by Ref.~\cite{mangano}.  These
assumptions reasonably reproduce the UA1 integrated cross-section (when
rapidity cuts are included)
in proton-antiproton collisions,
provided one uses an overall multiplicative
$K$-factor of roughly $K = 2.5$, a fact
which is consistent with many other analyses \cite{dawson2}.

The resulting average asymmetry in the integrated cross-section
in proton-proton collisions
is shown in Fig. 3(a) as a function of $p_T(min)$.
As expected, the value
at $p_T(min) = 0$, corresponding to the total cross-section,
gives a small
positive average asymmetry consistent with our observations above. For
values of $p_T(min) \geqnew 2M$, the spin asymmetry in
the dominant gluon fusion contribution is already approaching
its asymptotic
value of $-1$ so that the `analyzing power' of this reaction,
based on the
$2 \rightarrow 2$ processes, seems optimally large.  This reaction is
unique in that all contributing lowest order processes are characterized
by the same maximal partonic level asymmetry, at least at large
$p_T$.

For comparison, we
plot in Fig. 3(b)
the corresponding average $<\!\hat{a}_{LL}\!>$ for the $2 \rightarrow 2$
processes contributing to direct photon production, namely
$q+g \rightarrow q+\gamma$ and $q+\overline{q} \rightarrow g+\gamma$.
In this case also, increasing $p_T$ leads to $|y| \rightarrow 0$ where
$\hat{a}_{LL}$ is reduced compared to the larger values in the forward
and backward direction found at lower $p_T$. This helps to explain the
(slow) decrease in $<\hat{a}_{LL}>$ from to the $qg/gq$ contributions.
The partial cancellation between the $qg$ and $q\overline{q}$
contributions is also evident.

To estimate the effects of including  polarized gluon distributions,
we evaluate the observable asymmetry, $A_{LL}$, using a simplified
set of assumptions. We assume that the quark sea is unpolarized so that
there is no contribution from the $q \overline{q}$ diagrams.  For the
polarized gluon distribution, we use the ansatz,
$\Delta G(x,Q^2) = x^{\alpha} G(x,Q^2)$ which yields a value of
the integrated contribution to the proton spin
(at $Q^2_0 = 10\;\GeV^2$) of
\begin{equation}
\Delta G \approx 0.5, \;2.0, \;4.5 \;\;\;\;\;
 \mbox{for}\;\;\;\;\; \alpha = 1.0, \; 0.5, \; 0.25\;\;,
\end{equation}
where
\begin{displaymath}
\Delta G = \int_{0}^{1}\,dx\,\Delta G(x,Q^2_0)\,\,.
\end{displaymath}
This range of values spans many of
the models which have been discussed in the
literature.  The resulting asymmetries are shown in Fig.~4
and demonstrate the large variation in $A_{LL}$ which is possible.

This analysis based on the $2 \rightarrow 2$ subprocesses suggests
that not only could $b$-quark production at large $p_T$ be an
excellent process for the extraction of the polarized gluon distributions
(based on the large value of the average asymmetry) but that the
dramatic change in the sign of $A_{LL}$ implied by Fig.~3(a) would
constitute an interesting test of the spin-structure of the QCD
matrix elements.  All of this interesting spin structure is in a
region where perturbative QCD is applicable and at energies where
it has been thoroughly tested.  In addition, the energy is not so
high as to require the use of resummation techniques \cite{collins}
which are necessary when working at very low $x$.
The large $K$ factor necessary to
reproduce the data, however, is  reminder that radiative corrections
play an important role in the detailed structure of the observable
cross-section and we examine the possible effects of such
next-to-leading order effects in the next section.

\section{NLO  corrections and $2 \rightarrow 3$ processes}

The ability to confront or even to
extract unpolarized gluon distributions from $b$-quark production data
relies on the existence of a full set of next-to-leading order
 corrections
to heavy quark production \cite{dawson1,dawson2,smith}
which are by now standard.
Clearly any attempt to probe the {\bbf polarized}
gluon distributions will also eventually require a complete analysis of
the {\bbf spin-dependent} $O(\alpha_S^3)$ corrections as well.
In this section
we make a few comments on the likely effect that such corrections might
play in modifying the results found in Sec. II.  We do not attempt a
complete analysis of the spin-dependent radiative corrections.

The finite $O(\alpha_S^3)$ corrections remaining after the extraction of
divergent pieces to be associated with parton evolution will come in two
varieties: those involving vertex and box-diagram corrections to the
$2 \rightarrow 2$ processes and those arising from $2 \rightarrow 3$
subprocesses such as $g + g \rightarrow Q + \overline{Q} +g$.
The
first class of corrections, retaining as it does the same basic helicity
structure of the final state heavy quarks, will still yield a partonic
level spin-spin asymmetry approaching $-100\%$
at large transverse momentum,
i.e., $\hat{a}_{LL} \rightarrow -1$, when $\sqrt{\hat{s}}$
becomes large compared to the final state quark
masses.
 The approach
to this value could well be different but at large enough values of
$p_T$, say $p_T \geqnew 4M$,
we expect the same general features as shown in Fig. 3(a) to
appear. We cannot, for example, argue on general grounds
that the value of
$\hat{a}_{LL}$ near threshold will remain maximally large at $+1$.

Long before the complete NLO calculations were
available, it was argued \cite{kunszt}
that gluon fragmentation to heavy
quark pairs, via $2 \rightarrow 3$ subprocesses such as
\begin{eqnarray}
g + g \rightarrow g + g \rightarrow Q +\overline{Q} +g \nonumber \\
g + q \rightarrow g + q \rightarrow Q + \overline{Q} + q \nonumber\;\;,
\end{eqnarray}
is likely to be a large, even the
dominant source of heavy quarks at large transverse
momentum, i.e., for $p_T >> M$, yielding much, if not all, of the
needed $K$ factor.
The standard argument put forth is that at $90^{\circ}$ in the
parton-parton center-of-mass frame, gluon fusion is much more
likely to produce a gluon pair rather than quarks, i.e.,
\begin{equation}
\frac{\sum|{\cal M}(gg \rightarrow gg)|^2}
{\sum |{\cal M}(gg \rightarrow q\overline{q})|^2} \approx 200\;\;\;.
\end{equation}
The graphical structure of the $2 \rightarrow 3$ diagrams (dominated
by $t$- and $u$-channel gluon exchange) is very different from that
of the $2 \rightarrow 2$ diagrams (coming from $s$-channel annihilation
and heavy quark exchange) leading to different kinematics, and presumably,
a different spin structure.  In fact, the spin-spin asymmetries for the
underlying $2 \rightarrow 2$ processes shown in Fig. 1(b), namely
$g + g \rightarrow g + g$ and $q + g \rightarrow q + g$,
 are all
large and {\it positive} which leads to the disappointing possibility
that there might be substantial cancellations with  the spin-effects
present in the $g +g, q + \overline{q} \rightarrow Q + \overline{Q}$
processes.
Using the asymmetries in Fig. 1(b) (in the large $p_T$
limit we use the values at $y = \cos(\theta^*) \rightarrow 0$ of
roughly $0.6-0.8$)
and the observed
necessary $K$ factor of $K \approx 2.5$, we might estimate very crudely
that the average asymmetry could be reduced to something of the order
\begin{equation}
<\!\hat{a}_{LL}\!> \approx
\frac{1\cdot (-1) + (K-1) \cdot(0.6-0.8)}{1 + (K-1)}
\approx 0.08-(-0.04) \approx 0
\end{equation}
instead of $<\!\hat{a}_{LL}\!> = -1$.

To examine this possibility more quantitatively, we choose to model
the  effects of the $2 \rightarrow 3$
subprocesses by following one rather typical pre-radiative correction
analysis of such effects by Glover, Hagiwara, and Martin \cite{hagiwara}.
They `regularize' the $2 \rightarrow 3$ contributions by requiring that
\begin{equation}
\frac{p_T(Q\overline{Q})}{m(Q\overline{Q})} > \epsilon
\;\;\;\;\;\;\;\; , \;\;\;\;\;\;\;\;\;
\theta_{Qc}, \theta_{\overline{Q}c} > \delta
\end{equation}
where the notation $a + b \rightarrow c+ Q + \overline{Q}$ collectively
denotes the subprocesses $g+g \rightarrow g + Q + \overline{Q}$,
$q + g \rightarrow q + Q + \overline{Q}$, and $q + \overline{q}
\rightarrow g + Q + \overline{Q}$.  In Eqn. 18, $\epsilon$ and
$\delta$ are cutoffs which we take to be $\epsilon = 0.2$
and $\delta = 20^{\circ}$ respectively.  These conditions ensure
that the final light parton, (c), is hard and acollinear to the
initial parton directions while the angular cut excludes collinear
gluon emission from the final heavy quarks.  The remaining phase space
configurations are further restricted to three regions, described by
the following cuts:

\begin{flushleft}
`A': three-jet configuration
\end{flushleft}
\begin{eqnarray}
d(i,j) > 1 & & \mbox{for (i,j) =
$(Q,\overline{Q}), (c,Q), (c,\overline{Q})$}
\nonumber \\
p_T(i) > 10\,\GeV & & \mbox{for i = $Q,\overline{Q},c$}
\end{eqnarray}

\begin{flushleft}
`B': heavy quark excitation
\end{flushleft}
\begin{equation}
\mbox{Min$(p_T(Q),p_T(\overline{Q})) < 5\,\GeV$}
\end{equation}

\begin{flushleft}
`C': collinear $Q\overline{Q}$ production
\end{flushleft}
\begin{equation}
d(Q,\overline{Q})<1.
\end{equation}

In the equations above, $d(i,j)$ is the separation in the
pseudorapidity-azimuthal angle plane, namely,
\begin{equation}
d(i,j) = \left[(\Delta \eta_{ij})^2 + (\Delta \phi_{ij})^2 \right]^{1/2}
\;\;\;.
\end{equation}
The cut-off dependence of these prescriptions has been extensively
discussed in Ref.~\cite{hagiwara} where it is also found that the
massless limit of the exact $2 \rightarrow 3$ matrix elements
\cite{sexton} is an
excellent approximation at all but the very smallest values of
$p_T(min)$.   Making use of this simplified model of the important
contributions arising from $t$- and $u$-channel gluon exchange
processes, we plot in Fig. 5 the resulting integrated cross-sections
using both the $ 2 \rightarrow 2$ and
these regularized $ 2 \rightarrow 3$
subprocesses, here collected by initial parton combination ($gg$, $qg$,
and $q\overline{q}$.)  As expected, the latter diagrams dominate
the cross-section at large transverse momentum, in this case even roughly
reproducing the required $K \approx 2.5$ factor (which is somewhat
fortuitous).  These results are consistent with earlier calculations
which make similar assumptions \cite{halzen} but they do
seem to overestimate somewhat the
contribution from the $qg$ initial states compared to the exact
NLO calculation \cite{dawson2}.

Because the massless matrix elements are known to be a good approximation
in this case, we can also make use of the corresponding partonic
level asymmetries for the massless
$2 \rightarrow 3$ subprocesses which were
first evaluated in Ref.~\cite{rick3jet}.  We use these to evaluate
the resulting average asymmetry, $<\!\hat{a}_{LL}>$, now incorporating
all the $2 \rightarrow 2$ and $2 \rightarrow 3$ subprocesses. The
resulting average asymmetry in the integrated cross-section is
shown in Fig. 6 as a function of $p_T(min)$.  In each of the three
kinematic regions, the $q\overline{q}$ contributions are negligible
and in the $`A'$ configuration the individual contributions to the
average asymmetry from the $gg$ and $qg$ configurations are consistent
with existing results \cite{rick3jet}
on the spin asymmetries in three-jet events, namely small and positive
for the $qg$ configurations and small and negative for the $gg$ type.
In both the $`B'$ and $`C'$ classes, the spin-spin asymmetries arising
from the $gg$ and $qg$ contributions are relatively large and positive,
reflecting their connection to the underlying $gg \rightarrow gg$ and
$qg \rightarrow qg$ origin as seen in Fig. 1(b).
The intuitive and
qualitative estimates quoted above seem to be borne out in this
approximate calculation showing that there may well be
a large cancellation
in the spin-dependence.

We also note that the $2 \rightarrow 3$ configurations make
 a small negative
contribution to the asymmetry in the integrated cross-section
 at very small
values of $p_T(min)$, somewhat reducing the asymmetry in the total
cross-section.  This is, however, in the region where the massless
approximation for the matrix-elements and partonic asymmetries is worst
and, in fact, may likely give the wrong sign for $\hat{a}_{LL}$.
(Compare the purely massless
$2\rightarrow 2$ matrix elements and asymmetries to the exact ones which
would give a constant  asymmetry of $-1$ at threshold instead of $+1$.)
Combined
with the fact that the NLO corrections to the $2 \rightarrow 2$ diagrams
will likely give similar positive asymmetries near threshold,
 it seems that
the asymmetry in the total cross-section, or at least for $p_T(min) < M$,
could have a form quite similar to that obtained with
the LO $2 \rightarrow 2$ diagrams alone.  Once again, however, a complete
NLO spin-dependent calculation  will be required to verify these
conjectures.  One possible experimental difficulty with attempts at
extracting spin information from such low $p_T$ data is that the UA1
and CDF data both have lower bounds on $p_T(min)$ of roughly $5$ and
$10$ $\GeV$ respectively.  The total $b$-production cross-section that
UA1 quotes is then extracted from extrapolations of the $p_T$-dependent
data to $p_T(min) \rightarrow 0$ and to the full-rapidity interval
by using the QCD predictions for the rapidity variation. It is in just
this region that one would like real data to probe the spin-dependence.

We note that a very similar analysis could be applied to gluino
pair production using the $2 \rightarrow 2$ cross-sections and
asymmetries of Ref.~\cite{renard}, provided one could generalize
the results for $g + g \rightarrow \tilde{g} + \tilde{g} + g$ of
Ref.~\cite{gunion} to include spin-dependence.  In this case,
however, the use of any massless approximation for the matrix elements
would likely be unjustified due to the presumably large gluino mass.
Any realistic appraisal of the spin-dependence of gluino production
at supercollider energies will seemingly require a much more detailed
analysis than has been done to date.

Finally, it has been pointed out \cite{ji} that the {\bbf transverse}
spin-dependent quark and antiquark distributions (often called
`transversity') can be probed using the $q + \overline{q}
\rightarrow Q + \overline{Q}$ contribution to heavy quark production.
While this annihilation process does indeed have a large intrinsic
transverse polarization asymmetry, $\hat{a}_{TT}$, the overall
contribution of this subprocess to the total cross-section is never
large in $pp$ collisions at the energies we discuss, so that the
corresponding average asymmetry $<\hat{a}_{TT}>$, is never more than
$-5\%$.

\section{Conclusions}

While these results do not constitute a complete analysis of the
spin-dependence of $b$ quark production, they are suggestive of
some general trends which we believe will be a robust feature
of the full calculation.
 Inclusion of the polarized parton distributions
for a realistic evaluation of the observable asymmetries, $A_{LL}$, will
no doubt change the cancellations found above to some extent but we
certainly expect that the sizable cancellation found above between
LO  and NLO effects will persist in any complete analysis.  This
might well imply that $b$-quark production may not be the most sensitive
process from which to attempt to constrain or extract information on the
polarized gluon distribution, despite the highly suggestive LO results.

This cancellation is in itself, however,
a very interesting prediction which depends
sensitively on the spin structure of NLO QCD and could possibly
provide a useful
test of the matrix element structure of the strong interactions beyond
the leading order.  If the polarized gluon density were measured in
other reactions and found large, then the smallness of the observable
$A_{LL}$ in $b$-quark production
might then be attributable  to this cancellation and not to
the inherent smallness of $\Delta G(x,Q^2)$.

Next-to-leading corrections to jet or direct photon
production, for example,
for which the $O(\alpha_S^3)$ diagrams contain the same
kinematical structure would not be expected to yield such a dramatic
change in spin-dependence from the leading order result.  This is already
apparent from the NLO corrections for direct photon production which
have recently appeared \cite{conto2}.
The same conclusion is also suggested
by the fact that jet (direct photon) production is dominated at
leading order by spin-one gluon (spin-1/2 quark) exchange and this
feature is retained by NLO calculations and is confirmed by data
on jet-jet and jet-$\gamma$ angular distributions \cite{huth}.
The very different kinematical structures found in the $2 \rightarrow 2$
processes ($t$- and $u$-channel $Q$ exchange and $s$-channel production)
and in the $2 \rightarrow 3$ processes ($t$- and $u$-channel
gluon exchange)
could presumably be probed by similar $b$-quark pair angular
distributions
but this is experimentally difficult.  The spin-dependence of
 this process
might well provide a unique glimpse into the interplay between LO and NLO
corrections in QCD.  The only similar case evident so far in the
study of standard QCD processes at polarized colliders is the
case of double photon production which gets dominant contributions
from both $q+ \overline{q} \rightarrow \gamma + \gamma$ via Born diagrams
and from $g + g \rightarrow \gamma + \gamma$ via box-diagrams and
exhibits something of the same partial cancellation of spin-dependence
\cite{2gamma}.

Finally, and perhaps most importantly, these results
reinforce the notion that the
reliable extraction of spin-dependent parton distributions from data
obtained in polarized proton-proton collisions may  necessarily be
much more dependent on the availability of radiative corrections
than has been the case for unpolarized distributions in the past.

\section{Acknowledgments}

We thank J. D. Bjorken for emphasizing the importance of this process.
This work was supported in part by the National Science Foundation
under Grant No. PHY--9001744 (R.R.).
The  research of M.K. was supported in part
by grant No.~90-00342 from the United States-Israel
Binational Science Foundation (BSF), Jerusalem, Israel,
and by the Basic Research Foundation administered by the
Israel Academy of Sciences and Humanities.

\newpage
\single_space
\def\etal{{\em et al.}}
\def\PL{{ Phys. Lett.\ }}
\def\NP{{ Nucl. Phys.\ }}
\def\PR{{ Phys. Rev.\ }}
\def\PRL{{ Phys. Rev. Lett.\ }}

\newpage
\double_spacesp

{\Large Figure Captions}

\begin{itemize}

\item[Fig. \thinspace 1.] The partonic level spin-spin
asymmetry, $\hat{a}_{LL}$,
versus the center-of-mass scattering angle, $y = \cos(\theta^*)$ for
various partonic processes.\\
(a) The asymmetry for $g + g \rightarrow Q + \overline{Q}$
with the solid (dashed, dotdash, dotted) lines
corresponding to $\sqrt{\hat{s}}/2M = 1.1\,(1.5,\,2,5)$. \\
(b) The asymmetry $\hat{a}_{LL}$
for $g + g \rightarrow g + g$ (solid),
$q + g \rightarrow q + g (\gamma)$
(dotdashed) and $g + q \rightarrow q + g (\gamma)$ (dashed).

\item[Fig. \thinspace 2.] The spin-spin asymmetry,
 $\hat{A}_{LL} = \Delta \hat{\sigma}(\hat{s})/\hat{\sigma}(\hat{s})$ for
the integrated total cross-section for
$g + g \rightarrow Q + \overline{Q}$
versus $\sqrt{\hat{s}}/2M$.

\item[Fig. \thinspace 3.] (a) The average asymmetry,
$<\!\!\hat{a}_{LL}\!\!>$,
 for $b$-quark production in $pp$ collisions at $\sqrt{s} = 500\,\GeV$
using only $2 \rightarrow 2$ processes.
The total ($gg$, $q\overline{q}$)
contributions correspond to the solid (dashed, dotted) line.
(b) The same average asymmetry for direct
photon production using $2 \rightarrow 2$ subprocesses.  The total
($qg$, $q\overline{q}$) contributions correspond to the solid (dashed,
dotted) line.

\item[Fig. \thinspace 4.] The observable asymmetry $A_{LL}$  in the
integrated $b$-quark cross-section  ($\sigma(p_T > p_T(min))$),
versus $p_T(min) (\GeV)$.  A polarized gluon distribution of the
form $\Delta G(x,Q^2) = x^{\alpha} G(x,Q^2)$ is used with
$\alpha = 1.0,(\;0.5,\,0.25)$ for the dotted (dashed, solid) curves.

\item[Fig. \thinspace 5.] Integrated cross-section for $b$-quark
production, $\sigma(p_T > p_T(min)) (nb)$, versus $p_T(min) (\GeV)$,
in proton-proton collisions at $\sqrt{s} = 500\,\GeV$.
The upper (lower) dotted lines correspond to
$2\rightarrow2$ ($2 \rightarrow
3$) $gg$ processes.
The upper (lower) dotdash lines correspond to
$2\rightarrow2$ ($2 \rightarrow 3$) $q\overline{q}$ processes. The
dashed line corresponds to $2 \rightarrow 3$ $qg$ processes and the
solid line corresponds to the total.

\item[Fig. \thinspace 6.]  The average asymmetry $<\!\!\hat{a}_{LL}\!\!>$
 in the integrated
$b$-quark production cross-section versus $p_T(min) (\GeV)$. The solid
line is the result of  including
both the $2\rightarrow2$  and the
`regularized' $2\rightarrow 3$.
The dotted line is the total average asymmetry
when one considers $2 \rightarrow 2$ subprocesses only,
i.e. the result shown in Fig. 3(a).

\end{itemize}
\end{document}